\documentclass[preprint,review,12pt]{elsarticle}

\usepackage{amssymb}

\usepackage{hyperref}
\usepackage{amsmath}
\usepackage{cleveref}
\usepackage{booktabs}
\usepackage{caption,setspace}
\usepackage{tabularx}
\usepackage[export]{adjustbox}
\usepackage{tikz}
\usepackage{geometry}
\usepackage{wrapfig} 
\usepackage{flushend} 

\usepackage{helvet}  

\hypersetup{pdfauthor={Name}}

\makeatletter
\newcommand{\ssymbol}[1]{^{\@fnsymbol{#1}}}
\makeatother

\usepackage{array}
\newcolumntype{P}[1]{>{\centering\arraybackslash}p{#1}}

\usepackage{multirow}

\AtBeginDocument{
    \crefformat{equation}{#2Eq.~#1#3}
    \crefformat{figure}{#2Fig.~#1#3}
    \crefformat{table}{#2Tab.~#1#3}
}

\journal{arXiv}

\begin{document}
\begin{sloppypar}

\begin{frontmatter}

\title{CECT: Controllable Ensemble CNN and Transformer for COVID-19 Image Classification}


\author{Zhaoshan Liu\corref{}}
\ead{e0575844@u.nus.edu} 

\author{Lei Shen\corref{cor1}}
\ead{mpeshel@nus.edu.sg} 
\cortext[cor1]{Corresponding author}

\address{Department of Mechanical Engineering, National University of Singapore, 9 Engineering Drive 1, Singapore, 117575, Singapore}


\begin{abstract}
The COVID-19 pandemic has resulted in hundreds of million cases and numerous deaths worldwide. Here, we develop a novel classification network CECT by controllable ensemble convolutional neural network and transformer to provide a timely and accurate COVID-19 diagnosis. The CECT is composed of a parallel convolutional encoder block, an aggregate transposed-convolutional decoder block, and a windowed attention classification block. Each block captures features at different scales from 28 × 28 to 224 × 224 from the input, composing enriched and comprehensive information. Different from existing methods, our CECT can capture features at both multi-local and global scales without any sophisticated module design. Moreover, the contribution of local features at different scales can be controlled with the proposed ensemble coefficients. We evaluate CECT on two public COVID-19 datasets and it reaches the highest accuracy of 98.1\% in the intra-dataset evaluation, outperforming existing state-of-the-art methods. Moreover, the developed CECT achieves an accuracy of 90.9\% on the unseen dataset in the inter-dataset evaluation, showing extraordinary generalization ability. With remarkable feature capture ability and generalization ability, we believe CECT can be extended to other medical scenarios as a powerful diagnosis tool. Code is available at \href{https://github.com/NUS-Tim/CECT}{https://github.com/NUS-Tim/CECT}.
\end{abstract}


\begin{keyword}
Medical Image Classification \sep Convolutional Neural Network \sep Transformer \sep COVID-19



\end{keyword}

\end{frontmatter}



\section{Introduction}
\label{1}

COVID-19, an infectious disease induced by the novel coronavirus SARS-CoV-2 emerged towards the end of 2019. It is primarily transmitted through respiratory droplets and contact routes and the infection of COVID-19 can lead to symptoms such as fever and dry cough \citep{huang2020clinical}. According to the report, the rapid spreading of COVID-19 has resulted in over 600 million cases and 6 million deaths worldwide \citep{coronavirusmap}. To ensure a timely and accurate COVID-19 diagnosis, medical imaging, which generates visual representations of the interior body using modalities such as computed tomography and X-ray \citep{liu2022medical}, has been commonly leveraged during the diagnosis process. These imaging techniques can reveal lung abnormalities, providing valuable information for medical professionals for further medical intervention. Though diagnosing patients based on medical imaging is helpful, training and analysis procedures for human expertise are time-consuming. Thanks to the fast development of computer vision (CV), high-accuracy diagnosing results can be obtained instantly from well-trained CV models without the intervention of professional radiologists \citep{zhang2023mm}. The CV commits to extracting critical features from the visual inputs and is composed of several tasks, such as classification, segmentation, and denoising. Under the classification scene, the convolutional neural network (CNN) and transformer are two of the most widely used types of models.

CNN has dominated the CV field over the past several years and is capable of capturing spatial hierarchies of the input features with successive convolution operations \citep{yamashita2018convolutional}. A typical CNN consists of several types of layers, including convolution layer, dropout layer, normalization layer, etc. Compared with CNN, the transformer is an emerging role in the CV field initially proposed in the natural language processing field. The core component of the transformer is the self-attention mechanism or attention in short. The attention mechanism in a transformer network creates interdependencies among different positions within a single sequence, enabling the model to compute a context-aware representation of each position in the sequence \citep{vaswani2017attention}. The primary distinction between the CNN- and the transformer-based methods is that CNN tends to emphasize local features, whereas the transformer is more oriented toward global features. As the visual input may contain lesions of varying scales, it is crucial to enhance the ability of the model to capture features at both local and global scales. To take advantage of both CNN and transformer, the combination of CNN and transformer has developed rapidly. This kind of integration aids in capturing both local and global features and reducing the neglect of potentially diseased areas. Specifically, varying medical datasets may contain lesions of varying sizes. By capturing and understanding features across a broader range of scales, the hybrid model improves its ability to recognize and interpret features in unseen images, which may present lesions at varying scales. This extensive scale coverage enables the model to adapt to unseen images better, thereby significantly contributing to the overall performance and generalization capability.

Here, we develop a novel classification model CECT by Controllable Ensemble CNN and Transformer to improve the accuracy of COVID-19 diagnosis. The CECT is composed of a parallel convolutional encoder (PCE) block, an aggregate transposed-convolutional decoder (ATD) block, and a windowed attention classification (WAC) block. The PCE captures the features at \textit{multi-local} scales. The ATD decodes the captured features to the identical scale and sums them using proposed ensemble coefficients. The summed features are fed into the WAC to capture the \textit{global} features. Compared with existing approaches, CECT can extract features at both multi-local and global scales without complicated module design. Moreover, the contribution of local features at different scales can be controlled with the ensemble coefficients we proposed. Experimental results on the COVID-19 radiography dataset \citep{covid19radio, chowdhury2020can, rahman2021exploring} and the COVIDx CXR-3 dataset \citep{covidxcxr, wang2020covid} show the leadership of CECT. The highest accuracy of 98.1\% is achieved on the intra-dataset evaluation, outperforming state-of-the-art (SOTA) methods to a large extent. Moreover, CECT achieves a 90.9\% accuracy on the inter-dataset evaluation, demonstrating its extraordinary generalization ability. To sum up, our main contributions are:
\begin{itemize}
    \item We construct a novel classification model CECT to marry CNN and transformer to capture both multi-local and global features.
    \item The contribution across different local feature scales is controllable using proposed ensemble coefficients.
    \item Intra-dataset and inter-dataset evaluation results on two public datasets show that our CECT outperforms SOTA methods.
\end{itemize}

The rest of the paper is organized as follows. Section \hyperref[2]{2} "Related Work" illustrates the recent progress of CNN, transformer, and their combination. The corresponding applications in medical image classification are then introduced. Section \hyperref[3]{3} "Materials and Methods" discusses the datasets used and the implementation details of the proposed CECT, including architecture design, experimental setup, as well as evaluation metrics. The intra-dataset and inter-dataset evaluation results on two public COVID-19 datasets, together with the detailed analysis can be found in Section \hyperref[4]{4} "Results". Extensive ablation experiments are also conducted. We conclude our work and point out the future research perspectives in Section \hyperref[5]{5} "Conclusion".


\section{Related Work}
\label{2}

\subsection{CNN and Transformer}
\label{2.1}

CNN has established itself as one of the paramount models in the CV field. As of now, one of the prevalent models of CNNs is the ResNet \citep{he2016deep} developed by He et al. Instead of learning unreferenced functions, the ResNet reformulates the layers as learning residual functions regarding the layer inputs to ease the deeper network training. Based on the superior performance demonstrated by residual connections, Xie and colleagues \citep{xie2017aggregated} proposed an effective modular network architecture, which is constructed by repeating a building block that aggregates a set of transformations with the same topology. Recently, the evolution of CNN has seen a deceleration, partly attributed to the emergence of the transformer. To this end, numerous studies have sought to reaffirm the significance of CNN in the CV field. Among these efforts, the ConvNeXt \citep{liu2022convnet} developed by Liu et al. stands as one of the most accomplished. The ConvNeXt gradually "modernizes" the standard ResNet toward the architecture of the transformer \citep{dosovitskiy2020image}, thereby elucidating the pivotal components that contribute to the performance gap. In concert with ConvNeXt, the ConvFormer \citep{yu2022metaformer} that instantiates the token mixer as separable depthwise convolutions also shows SOTA performance. Additionally, Ding and colleagues \citep{ding2022scaling} introduced RepLKNet, which utilizes convolution kernels as large as 31 × 31 and significantly narrows the performance gap between the CNN and transformer. In 2023, Yu et al. \citep{yu2023inceptionnext} developed InceptionNeXt with inception depthwise convolution, which decomposes large-kernel depthwise convolution along channel dimension into varying parallel branches. The transformer has ascended to the forefront research field and has demonstrated superior performance over CNN. Inspired by its superior performance, a significant amount of subsequent research has been recently undertaken. For instance, Liu et al. \citep{liu2021swin} developed the swin transformer (SwT) with hierarchical feature maps. In SwT, the attention is computed within each local window, and the window connections are modeled by shifting the window partitioning after computing the attention. With such designs, the SwT can both model at various scales with high flexibility and reduce the computation complexity from quadratic to linear. Touvron and colleagues \citep{touvron2021training} proposed a deep-narrow T2T-ViT model, which incorporates a layerwise Tokens-to-Token transformation. The transformation enables the progressive structuring of images into tokens by iteratively combining adjacent tokens into one. Additionally, Han et al. \citep{han2021transformer} developed a TNT approach, in which local patches are regarded as “visual sentences.” These sentences are then divided into smaller “visual words,” and features from both words and sentences are aggregated to enhance the representational ability. 

In pursuit of leveraging the merits of both CNN and transformer, a significant amount of work has been dedicated to combining CNN and transformer in different ways. Among these, the most common way is to marry CNN and transformer either in tandem or parallel. For example, Duong et al. \citep{duong2021detection} developed a ViT-Eff method, in which input images are fed into EfficientNet \citep{tan2019efficientnet} and the extracted feature maps are then fed to the transformer and prediction head. Dong and colleagues \citep{dong2022cswin} proposed a CSWin Transformer, in which the convolutional token embedding is applied to the input ahead of feeding to consequent CSwin Transformer blocks with cross-shaped window attention. Tu et al. \citep{tu2022maxvit} proposed a MaxViT with a scalable attention mechanism termed multi-axis attention, in which the input is fed into the convolution stem followed by the transformer blocks. The multi-axis attention comprises blocked local and dilated global attention, enabling spatial interactions between global and local elements for inputs of any resolution while maintaining linear complexity. Comparable work is seen in CrossViT \citep{chen2021crossvit} proposed by Chen et al., in which convolution is implemented to achieve patch embedding. The CrossViT comprises a dual-branch transformer and combines image patches at different sizes to produce stronger image features. Besides, a cross-attention-based token fusion module is developed. For each branch, a single token is utilized as a query for information exchange across branches. Instead of tandem or parallel combinations, certain studies focus on fusing CNN and transformer in a more intrinsic way with reconfigured modules. A case in point is the CMT \citep{guo2022cmt} developed by Guo et al. The CMT follows the design of ResNet and is composed of several reimagined blocks, such as the local perception unit, lightweight multi-head attention, and inverted residual feed-forward network. Besides, Xu and colleagues \citep{xu2021co} proposed a CoaT method endowed with co-scale and conv-attentional mechanisms. The co-scale mechanisms preserve the cohesiveness of the encoder branches of the transformer while the conv-attentional mechanism achieves relative position embedding. Pan et al. \citep{pan2023slide} proposed a Slide-Transformer model with a novel slide attention module derived by substituting shifting operations with designed depthwise convolutions. It can not only impose local inductive bias like local attention but also maintain high flexibility and efficiency.

\subsection{Medical Image Classification}
\label{2.2}

Both CNN and transformer have been extensively utilized in medical image classification and achieved superior performance. For CNN, Abbas et al. \citep{abbas2021classification} developed a DeTraC model to detect COVID-19 from chest X-ray images. The DeTraC consists of three stages, which are decompose, transfer, and compose. Liu and colleagues \citep{liu2023gsda} leveraged a variety of CNN models to classify breast cancer images, in which the generative adversarial network was implemented for data augmentation. Wang et al. \citep{wang2020covid} developed a COVID-Net with high architectural diversity as well as selective long-range connectivity. In 2023, Hasan and colleagues \citep{hasan2023fp} introduced a novel fuzzy pooling-based model termed FP-CN, which leverages fuzzy pooling to extract representative features from visual images. Furthermore, the model employs Shapley Additive Explanations to elucidate the decision-making processes in its intermediate layers. Additionally, Aurna et al. \citep{aurna2022classification} developed a two-stage feature ensemble CNN to precisely classify brain tumors, in which the best-performing models are concatenated in two stages for feature extraction. Subsequently, Principal Component Analysis is employed to select the most significant features. Regarding the transformer, Krishnan et al. \citep{krishnan2021vision} employed the transformer for COVID-19 diagnosis. Aladhadh and colleagues \citep{aladhadh2022effective} utilized the transformer to classify skin images for skin cancer diagnosis. Images are processed using the data augmentation technique including rotation, flip, contrast, scaling, etc. Ayana and colleagues \citep{ayana2023vision} leveraged varying SOTA transformer models to classify breast mass mammograms and compared their performance. Gokhale et al. \citep{gokhale2023genevit} introduced a novel GeneViT model to classify cancerous gene expression. Specifically, the stacked autoencoder is implemented to reduce the dimensionality and the improved DeepInsight algorithm is utilized to convert the data into image format ahead of feeding to the transformer. Gheflati and colleagues \citep{gheflati2022vision} utilized the transformer to classify breast ultrasound images and adopted a weighted cross-entropy loss function to address the issue of the imbalanced dataset.

However, as the CNN- and transformer-based methods focus more on capturing local and global features respectively, marrying these two methods can further enrich the local and global features extracted. The recent methods of integrating CNN and transformer can be primarily categorized in two-fold. Firstly, the CNN and transformer are ensembled in tandem or parallel. For example, Tummala et al. \citep{tummala2022breast} ensembled a variety of versions of the swin transformer and the final model output is obtained by averaging the predicted softmax vectors of individual models. Dai and colleagues \citep{dai2021transmed} proposed a TransMed method, in which ResNet is leveraged to capture features, followed by the transformer. Liang \citep{liang2023light} et al. developed a hybrid light-weight model architecture based on CNN and transformer, in which convolution layers and MobileViT blocks \citep{mehta2021mobilevit} are connected in tandem. Secondly, the CNN and transformer are married more intrinsically with complicated redesigned modules. A case in this is the MedViT proposed in 2023 \citep{manzari2023medvit} with the developed locally feed-forward network within the local transformer block. An alternative can be the EDCA-Net developed by Zhu et al. \citep{zhu2023evolutionary}, consisting of the densely connected attention module with multiple densely connected channel-attentional feature units. In addition, Jiang et al. \citep{jiang2022mxt} developed an MXT architecture consisting of five stages. In MXT, the downsampling spatial reduction attention reduces resource usage while the multi-layer overlap patch tokenizes the images. Moreover, multi-label attention as well as the class token transformer block are incorporated, thereby providing a more effective procedure for multi-label scenarios. The convolution layers are integrated into the downsampling spatial reduction transformer block. Leveraging both local and global features captured, these methods demonstrate excellent performance. 

To further enhance the feature capture and generalization ability, we propose a straightforward yet effective CECT approach. Different from existing models, CECT can extract features at both \textit{multi-local} and \textit{global} scales without complicated module design. Moreover, the contribution of local features at varying scales is controllable through our proposed ensemble coefficients. Compared with tandem or parallel combinations, our CECT comprises three CNN-based branches designed to identify features at multi-local scales instead of a specific local scale. Contrasting with the approaches with sophisticated module design, our CECT exhibits enhanced effectiveness and generalization ability with straightforward yet effective architecture.


\section{Materials and Methods}
\label{3}

\subsection{Datasets}
\label{3.1}

\begin{figure*}
	\centering
	  \includegraphics[width=0.7\textwidth]{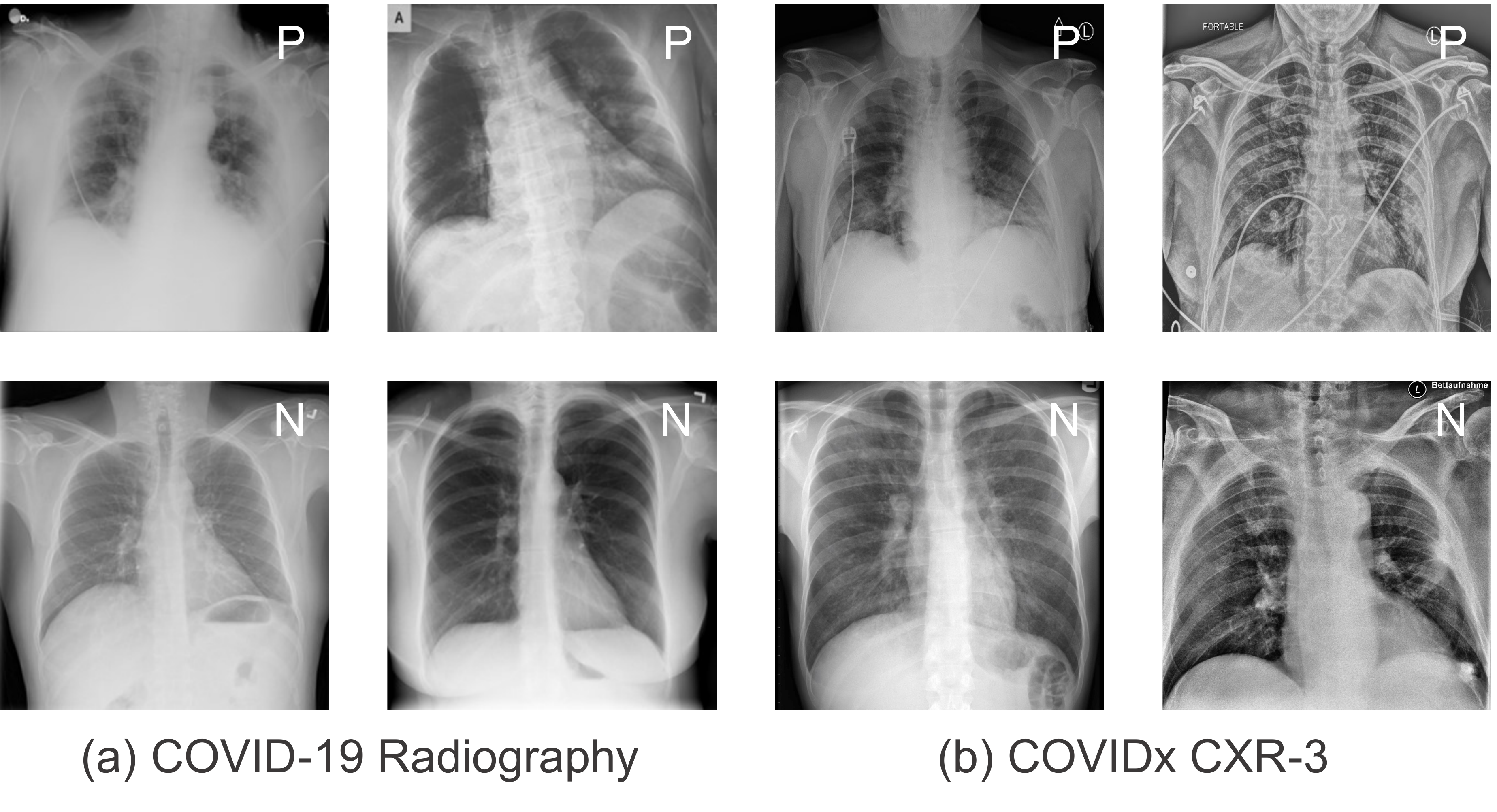}
	  \caption{Illustrative images from the two datasets. P stands COVID-positive while N presents COVID-negative.}
	  \label{fig1}
\end{figure*}

We evaluate CECT on the COVID-19 radiography dataset and the COVIDx CXR-3 dataset. The COVID-19 radiography dataset is collected by researchers from Qatar University and the University of Dhaka together with their collaborators. There are 3616 COVID-positive and 10192 COVID-negative images in the dataset. The COVIDx CXR-3 dataset is a chest X-ray dataset collected from 16648 patients at different locations. There are a total of 30386 images in the COVIDx CXR-3 dataset composed of two parts. The first part consists of 29986 images for training and validation, with 15994 labeled as positive and 13992 labeled as negative. The second part contains 400 images for testing, evenly split with 200 positive and 200 negative images. Illustration images from the two datasets can be found in \cref{fig1}.

\subsection{Model Architecture}
\label{3.2}

\begin{figure*}
	\centering
	  \includegraphics[width=0.8\textwidth]{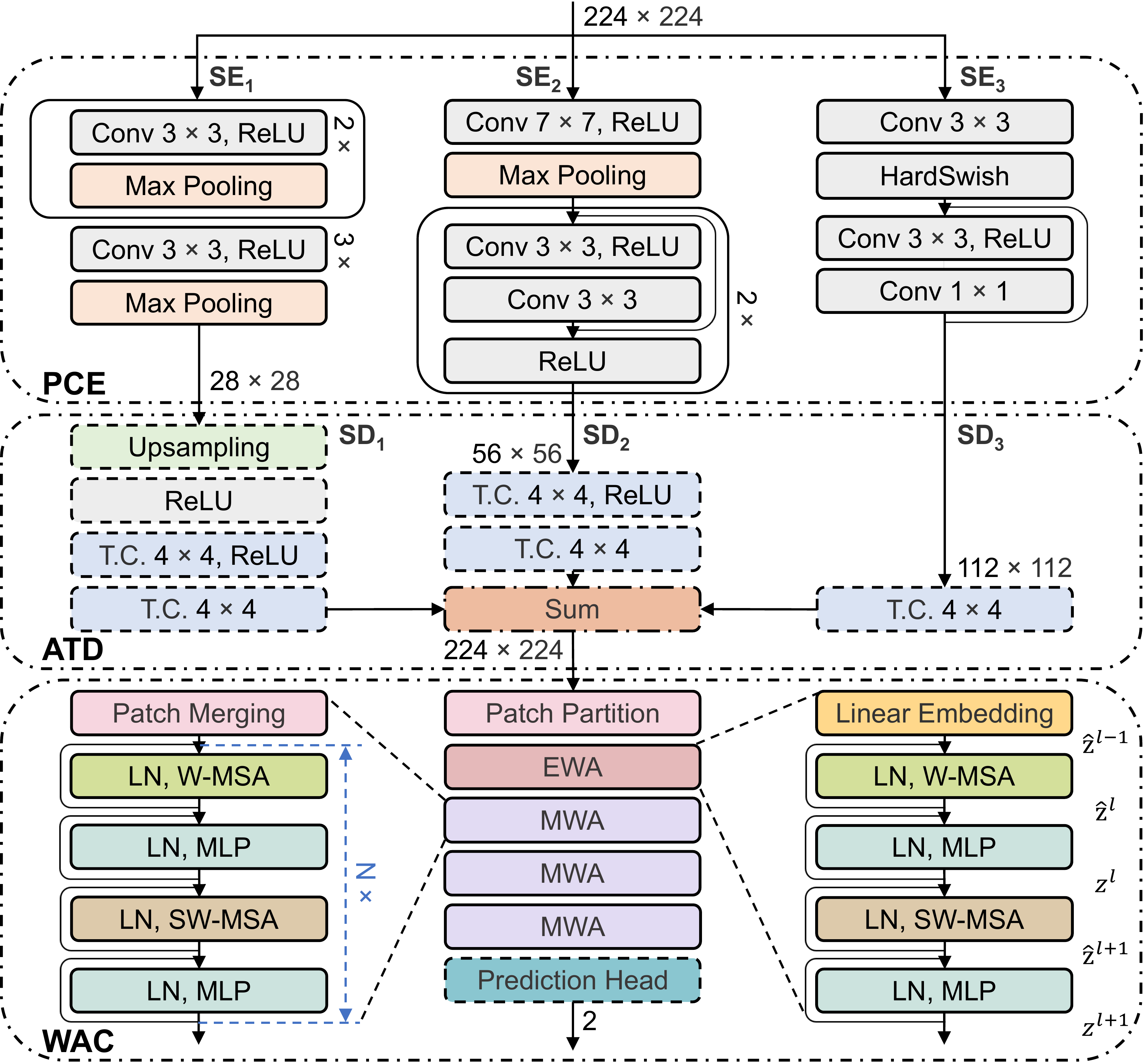}
	  \caption{Detailed architecture of proposed CECT. CECT consists of three blocks including the parallel convolutional encoder (PCE) block, aggregate transposed-convolutional decoder (ATD) block, and windowed attention classification (WAC) block. PCE consistes three sub-encoders, $\mathrm{SE}_1$, $\mathrm{SE}_2$, and $\mathrm{SE}_3$, while ATD constitutes three sub-decoders, $\mathrm{SD}_1$, $\mathrm{SD}_2$, and $\mathrm{SD}_3$. WAC contains multiple embedding window attention (EWA) and merging window attention (MWA) blocks, which consist of the linear embedding layer and patch merging layer ahead of the multi-head attention module, respectively. The attention module consists of sequential regular windowing attention (W-MSA) and shifted windowing attention (SW-MSA). The MWA blocks constitute $N$ attention modules, in which $N=\{1, 3, 1\}$ in the CECT setup. MLP represents the multilayer perceptron module. LN denotes layer normalization. $\hat{\mathbf{z}}^l$ and $\mathbf{z}^l$ stand the output features of the W-MSA and the MLP inside block $l$, respectively. T.C. represents the transposed convolutional layer. The solid and dotted lines denote the network components with and without transfer learning, respectively.}
	  \label{fig2}
\end{figure*}

The detailed architecture of CECT can be found in \cref{fig2}. CECT consists of three blocks, the PCE, ATD, and WAC. The PCE constitutes three sub-encoders, $\mathrm{SE}_1$, $\mathrm{SE}_2$, and $\mathrm{SE}_3$, which leverage part of VGGNet \citep{simonyan2015very}, ResNet, and MobileNet \citep{howard2019searching} as the backbone, respectively. To keep consistent scaling with the 224 × 224 scale global features, the output scales of $\mathrm{SE}_1$, $\mathrm{SE}_2$, and $\mathrm{SE}_3$ are organized as 28 × 28, 56 × 56, and 112 × 112, respectively. With multi-scale outputs, the local features at 28 × 28, 56 × 56, and 112 × 112 scales can be captured. The ATD comprises three sub-decoders $\mathrm{SD}_1$, $\mathrm{SD}_2$, and $\mathrm{SD}_3$, as well as a sum operation at the feature map level. We devised the sub-decoders from scratch with an output dimension of 224 × 224. Besides, the input dimensions of $\mathrm{SD}_1$, $\mathrm{SD}_2$, and $\mathrm{SD}_3$ are formulated to match those of the output dimensions of $\mathrm{SE}_1$, $\mathrm{SE}_2$, and $\mathrm{SE}_3$, respectively. The output feature maps are then summed using the proposed ensemble coefficients. The WAC is developed based on SwT and mainly consists of embedding window attention (EWA) block and merging window attention (MWA) block. The WAC takes the 224 × 224 summed feature maps as the input to capture features at the global scale. Through marrying PCE, ATD, and WAC, the CECT can capture both \textit{multi-local} and \textit{global} features from 28 × 28 to 224 × 224 scales. With the integration of multi-local and global features, the CECT can not only further decrease the probability of neglecting potential lesion areas but also exhibit enhanced generalization ability across diverse datasets.

In PCE, The $\mathrm{SE}_1$, $\mathrm{SE}_2$, and $\mathrm{SE}_3$ are intercepted from the top layers of the VGGNet, ResNet, and MobileNet, respectively. Specifically, the $\mathrm{SE}_1$ is extracted from the top 17 child layers of the VGGNet, while the $\mathrm{SE}_2$ and $\mathrm{SE}_3$ are intercepted from the top 5 and 2 child layers, from ResNet and MobileNet, respectively. In ATD, the $\mathrm{SD}_1$, $\mathrm{SD}_2$ and $\mathrm{SD}_3$ consists of the upsampling layer and various numbers of the 4 × 4 transposed convolutional (T.C.) layer. RELU is implemented as the activation function. The sum operation at the feature map level is formulated based on the developed coefficients $\alpha$, $\beta$, $\gamma$, in which $\alpha, \beta, \gamma \in[0,1]$ and $\alpha + \beta + \gamma = 1$. Each coefficient weights the contribution of the corresponding sub-decoders, thereby regulating the integration of features extracted at different scales. Through a rule-based search for the optimal combination of the coefficients, the CECT can adapt its feature extraction strategy to accommodate the characteristics of diverse datasets. This coefficient-tuning process is crucial for balancing the influence of multi-scale local features and ensuring generalization ability. Given $\mathcal{F}_{SD_1}$, $\mathcal{F}_{SD_2}$, and $\mathcal{F}_{SD_3}$ as the output feature maps of $\mathrm{SD}_1$, $\mathrm{SD}_2$, and $\mathrm{SD}_3$, the output $y$ of ATD can be formulated as:
\begin{equation}
y=\alpha \mathcal{F}_{SD_1}+\beta \mathcal{F}_{SD_2}+\gamma \mathcal{F}_{SD_3},
\label{eq1}
\end{equation}
The WAC consists of four stages, mainly with the EWA block and MWA blocks. Both EWA and MWA consist of the multi-head attention module with regular windowing attention (W-MSA) and shifted windowing attention (SW-MSA), in which the EWA and MWA process the input with the patch embedding layer and patch merging layer, respectively. The multilayer perceptron (MLP) module consists of two layers and a GELU \citep{hendrycks2016gaussian} activation function in between. The prediction head is composed of a linear layer with two output nodes. Given $\hat{\mathbf{z}}^l$ and $\mathbf{z}^l$ illustrate the output features of the W-MSA and the MLP inside the block $l$, respectively, the calculation of the attention with relative position bias in EWA and MWA blocks can be formulated as:
\begin{equation}
\hat{\mathbf{z}}^l=\operatorname{W-MSA}\left(\mathrm{LN}\left(\mathbf{z}^{l-1}\right)\right)+\mathbf{z}^{l-1},
\label{eq2}
\end{equation}
\begin{equation}
\mathbf{z}^l=\operatorname{MLP}\left(\operatorname{LN}\left(\hat{\mathbf{z}}^l\right)\right)+\hat{\mathbf{z}}^l,
\label{eq3}
\end{equation}
\begin{equation}
\hat{\mathbf{z}}^{l+1}=\operatorname{SW-MSA}\left(\mathrm{LN}\left(\mathbf{z}^l\right)\right)+\mathbf{z}^l,
\label{eq4}
\end{equation}
\begin{equation}
\mathbf{z}^{l+1}=\operatorname{MLP}\left(\operatorname{LN}\left(\hat{\mathbf{z}}^{l+1}\right)\right)+\hat{\mathbf{z}}^{l+1},
\label{eq5}
\end{equation}

\subsection{Experimental Setup}
\label{3.3}

\begin{table}
\centering
\caption{The number of images in each subset of the two datasets.}
\resizebox{0.7\linewidth}{!}{
\begin{tabular*}{356pt}{cccccc}
\toprule
    Dataset                               & Category   & Positive & Negative & Total \\
\midrule
    \multirow{4}{*}{COVID-19 radiography} & Training   & 2892     & 8154     & 11046 \\
                                          & Validation & 362      & 1019     & 1381  \\
                                          & Test       & 362      & 1019     & 1381  \\
                                          & Total      & 3616     & 10192    & 13808 \\
\midrule
    \multirow{4}{*}{COVIDx CXR-3}         & Training   & 12795    & 11194    & 23989 \\
                                          & Validation & 3199     & 2798     & 5997  \\
                                          & Test       & 200      & 200      & 400   \\
                                          & Total      & 16194    & 14192    & 30386 \\
\bottomrule
\end{tabular*}
}
\label{tab1}
\end{table}

In our setup, we randomly divide each dataset into different subsets. For the COVID-19 radiography dataset, we divided it into a training subset, a validation subset, and a test subset with a ratio of 8:1:1. As for the COVIDx CXR-3 dataset, we divided it into a training subset and a validation subset with a ratio of 8:2 as the images for testing are provided separately. We maintain the same ratio of positive to negative examples in the training, validation, and test subsets to ensure class balance across each. Details of the dataset partition results can be found in \cref{tab1}. Images are resized and cropped to the resolution of 224 × 224 and normalized. Data augmentations including random resize, crop, and horizontal flip are implemented for the training subset for generalization ability consideration. We train various models for 20 epochs with an initial learning rate of 0.003. As the training proceeds, the learning rate is decreased with a factor of 0.5 and patience of 5. The training batch size is 64. We leverage the cross-entropy loss as the loss function and set the optimizer as Adam. Transfer learning is implemented for the PCE and WAC, excluding the prediction head of WAC. Specifically, these network components are pre-trained on the ImageNet \citep{deng2009imagenet} ahead of training. We show the network component with and without transfer learning in solid and dotted lines in \cref{fig2}, respectively. To manage the computation consumption within a reasonable boundary, we stipulate the value of $\alpha$, $\beta$, and $\gamma$ selected from \{0.1, 0.2, $0.\dot{3}$, 0.6, 0.8\}, in which $0.\dot{3}$ represents $\frac{1}{3}$. Moreover, two or more coefficients share the same value. Under such setup, seven groups of $\alpha$, $\beta$, and $\gamma$ are generated, as shown in \cref{tab2} and \cref{tab3}. We perform our experiments using NVIDIA RTX A6000 and Intel Xeon 3204 on Pytorch 1.13.1. Detailed libraries can be accessed from our GitHub repository. 

\subsection{Evaluation Metrics}
\label{3.4}

We leverage various evaluation metrics to assess the model performance, including accuracy (ACC), negative predictive value (NPV), positive predictive value (PPV), sensitivity (SEN), specificity (SPE), and F-1 score (FOS). The evaluation metrics are calculated based on true positive (TP), false negative (FN), false positive (FP), and true negative (TN), corresponding to sub-blocks of the confusion matrix in our setup. TP and TN denote the counts of correctly predicted positive and negative instances, respectively, while FP and FN represent counts of incorrect predictions for the respective categories. Given TP, TN, FP, and FN, the evaluation metrics can be calculated using:
\begin{equation}
A C C=\frac{T P+T N}{T P+T N+F P+F N},
\label{eq6}
\end{equation}
\begin{equation}
N P V=\frac{T N}{T N+F N},
\label{eq7}
\end{equation}
\begin{equation}
P P V=\frac{T P}{T P+F P},
\label{eq8}
\end{equation}
\begin{equation}
S E N=\frac{T P}{T P+F N},
\label{eq9}
\end{equation}
\begin{equation}
S P E=\frac{T N}{T N+F P},
\label{eq10}
\end{equation}
\begin{equation}
F O S=\frac{2 T P}{2 T P+F N+F P},
\label{eq11}
\end{equation}


\section{Results}
\label{4}

\subsection{Ensemble Coefficients}
\label{4.1}

\begin{table}[ht]
\centering
\caption{The performance of the CECT under various groups of $\alpha$, $\beta$, and $\gamma$ on the COVID-19 radiography dataset. $0.\dot{3}$ represents $\frac{1}{3}$. The best results are in bold.}
\resizebox{0.64\linewidth}{!}{
\begin{tabular*}{362pt}{ccccccccc}
\toprule
    $\alpha$    & $\beta$     & $\gamma$    & ACC             & NPV             & PPV             & SEN             & SPE             & FOS             \\
\midrule
    0.8         & 0.1         & 0.1         & 98.0\%          & 98.8\%          & 95.9\%          & 96.7\%          & 98.5\%          & 96.3\%          \\
    0.6         & 0.2         & 0.2         & 97.6\%          & \textbf{99.2\%} & 93.4\%          & \textbf{97.8\%} & 97.5\%          & 95.5\%          \\
    0.1         & 0.8         & 0.1         & 97.8\%          & 98.6\%          & 95.3\%          & 96.1\%          & 98.3\%          & 95.7\%          \\
    0.2         & 0.6         & 0.2         & 98.0\%          & 98.2\%          & \textbf{97.5\%} & 95.0\%          & \textbf{99.1\%} & 96.2\%          \\
    0.1         & 0.1         & 0.8         & 97.5\%          & 98.6\%          & 94.6\%          & 96.1\%          & 98.0\%          & 95.3\%          \\
    0.2         & 0.2         & 0.6         & 98.0\%          & 98.8\%          & 95.6\%          & 96.7\%          & 98.4\%          & 96.2\%          \\
    $0.\dot{3}$ & $0.\dot{3}$ & $0.\dot{3}$ & \textbf{98.1\%} & 98.6\%          & 96.7\%          & 96.1\%          & 98.8\%          & \textbf{96.4\%} \\
\bottomrule
\end{tabular*}
}
\label{tab2}
\end{table}

\begin{table}[ht]
\centering
\caption{The performance of the CECT under various groups of $\alpha$, $\beta$, and $\gamma$ on the COVIDx CXR-3 dataset.}
\resizebox{0.64\linewidth}{!}{
\begin{tabular*}{362pt}{ccccccccc}
\toprule
    $\alpha$    & $\beta$     & $\gamma$    & ACC             & NPV             & PPV             & SEN             & SPE             & FOS             \\
\midrule
    0.8         & 0.1         & 0.1         & 93.2\%          & 88.8\%          & \textbf{98.9\%} & 87.5\%          & 99.0\%          & 92.8\%          \\
    0.6         & 0.2         & 0.2         & 87.2\%          & 80.2\%          & 98.7\%          & 75.5\%          & 99.0\%          & 85.6\%          \\
    0.1         & 0.8         & 0.1         & 92.7\%          & 88.3\%          & 98.3\%          & 87.0\%          & 98.5\%          & 92.3\%          \\
    0.2         & 0.6         & 0.2         & 90.0\%          & 83.9\%          & 98.8\%          & 81.0\%          & 99.0\%          & 89.0\%          \\
    0.1         & 0.1         & 0.8         & 89.5\%          & 82.9\%          & 99.4\%          & 79.5\%          & \textbf{99.5\%} & 88.3\%          \\
    0.2         & 0.2         & 0.6         & 89.0\%          & 82.5\%          & 98.7\%          & 79.0\%          & 99.0\%          & 87.8\%          \\
    $0.\dot{3}$ & $0.\dot{3}$ & $0.\dot{3}$ & \textbf{97.2\%} & \textbf{96.1\%} & 98.5\%          & \textbf{96.0\%} & 98.5\%          & \textbf{97.2\%} \\
\bottomrule
\end{tabular*}
}
\label{tab3}
\end{table}

We perform intensive experiments with CECT using seven groups of $\alpha$, $\beta$, and $\gamma$ on the COVID-19 radiography dataset and COVIDx CXR-3 dataset in \cref{tab2} and \cref{tab3}, respectively. We observed that the coefficient group with values of $0.\dot{3}$, $0.\dot{3}$, and $0.\dot{3}$ yields the optimal performance compared with the remaining setups. Form \cref{tab2}, it is found that values of $\alpha$, $\beta$, and $\gamma$ have no discernible impact on the model performance and the evaluation metrics fluctuate between 93.4\% and 99.2\%. When $\alpha$, $\beta$, and $\gamma$ all equal $0.\dot{3}$, highest ACC and FOS of 98.1\% and 96.4\% are observed. When $\alpha$, $\beta$, and $\gamma$ equals 0.2, 0.6, and 0.2, the highest PPV and SPE of 97.5\% and 99.1\% are achieved, respectively. The highest NPV and SEN of 99.2\% and 97.8\% are observed when $\alpha$, $\beta$, and $\gamma$ equals 0.6, 0.2, and 0.2. We choose the coefficient group of $0.\dot{3}$, $0.\dot{3}$, and $0.\dot{3}$ for performance comparison in Section \hyperref[4.2]{4.2} as it shows the highest ACC. Regarding the \cref{tab3}, it can be found that the $\alpha$, $\beta$, and $\gamma$ show a significant impact on the model performance as the evaluation metrics vary from 75.5\% to 99.5\%. When $\alpha$, $\beta$, and $\gamma$ all equal $0.\dot{3}$, the highest ACC, NPV, SEN, and FOS are observed, reaching 97.2\%, 96.1\%, 96.0\%, and 97.2\%, respectively. The highest SPE of 99.5\% is achieved when $\alpha$, $\beta$, and $\gamma$ equals 0.1, 0.1, and 0.8. When $\alpha$, $\beta$, and $\gamma$ equals 0.8, 0.1, and 0.1, the highest PPV of 98.9\% is observed. Significant variations in model performance among different coefficient groups can potentially be attributed to the data distribution. In case of encountering inconsistent data distribution, features at different scales all count. To this end, capturing features equally at all scales can improve the overall performance to a large extent. We choose the coefficient group of $0.\dot{3}$, $0.\dot{3}$, and $0.\dot{3}$ for performance comparison in Section \hyperref[4.2]{4.2} due to the overall performance leadership.

\subsection{Comparison and Analysis}
\label{4.2}

\begin{table}[ht]
\centering
\caption{Performance comparison between the CECT and SOTA models on the COVID-19 radiography dataset. Pure methods are based solely on CNN or transformer, while hybrid approaches integrate both.}
\resizebox{0.84\linewidth}{!}{
\begin{tabular*}{504pt}{ccccccccccc}
\toprule
    Category  & Model                                                  & Venue     & ACC             & NPV             & PPV             & SEN             & SPE             & FOS             \\
\midrule
    \multirow{8}{*}{Pure}   & ConvNeXt \citep{liu2022convnet}          & CVPR 22'  & 96.2\%          & 96.6\%          & 94.8\%          & 90.3\%          & 98.2\%          & 92.5\%          \\
                            & RegNetX \citep{radosavovic2020designing} & CVPR 20'  & 95.9\%          & 97.3\%          & 92.0\%          & 92.5\%          & 97.2\%          & 92.3\%          \\
                            & GhostNet \citep{han2020ghostnet}         & CVPR 20'  & 97.2\%          & 96.9\%          & \textbf{98.2\%} & 91.2\%          & \textbf{99.4\%} & 94.6\%          \\
                            & CSPNet \citep{wang2020cspnet}            & CVPRW 20' & 92.7\%          & 94.6\%          & 87.0\%          & 84.8\%          & 95.5\%          & 85.9\%          \\
                            & SKNet \citep{li2019selective}            & CVPR 19'  & 95.3\%          & 97.4\%          & 89.6\%          & 92.8\%          & 96.2\%          & 91.2\%          \\
                            & VoVNet \citep{lee2019energy}             & CVPRW 19' & 95.7\%          & 96.3\%          & 93.9\%          & 89.5\%          & 97.9\%          & 91.7\%          \\
                            & DeiT \citep{touvron2021training}         & ICML 21'  & 95.4\%          & 95.1\%          & 96.3\%          & 85.6\%          & 98.8\%          & 90.6\%          \\
                            & ConViT \citep{d2021convit}               & ICML 21'  & 92.7\%          & 92.3\%          & 94.2\%          & 76.8\%          & 98.3\%          & 84.6\%          \\
\midrule    
    \multirow{8}{*}{Hybrid} & MaxViT \citep{tu2022maxvit}              & ECCV 22'  & 92.0\%          & 91.9\%          & 92.3\%          & 75.7\%          & 97.7\%          & 83.2\%          \\
                            & CrossViT \citep{chen2021crossvit}        & ICCV 21'  & 79.7\%          & 84.2\%          & 63.6\%          & 53.0\%          & 89.2\%          & 57.8\%          \\
                            & LeViT \citep{graham2021levit}            & ICCV 21'  & 90.3\%          & 90.2\%          & 90.7\%          & 70.2\%          & 97.4\%          & 79.1\%          \\
                            & CoaT \citep{xu2021co}                    & ICCV 21'  & 92.8\%          & 94.3\%          & 88.1\%          & 83.7\%          & 96.0\%          & 85.8\%          \\
                            & MobileViT \citep{mehta2021mobilevit}     & ICLR 22'  & 82.8\%          & 84.5\%          & 74.8\%          & 51.7\%          & 93.8\%          & 61.1\%          \\
                            & CMT \citep{guo2022cmt}                   & CVPR 22'  & 94.6\%          & 94.9\%          & 93.9\%          & 85.1\%          & 98.0\%          & 89.3\%          \\
                            & MedViT \citep{manzari2023medvit}         & CIBM 23'  & 95.6\%          & 95.5\%          & 95.7\%          & 87.0\%          & 98.6\%          & 91.2\%          \\
                            & \textbf{CECT(Ours)}                      & -         & \textbf{98.1\%} & \textbf{98.6\%} & 96.7\%          & \textbf{96.1\%} & 98.8\%          & \textbf{96.4\%} \\
\bottomrule
\end{tabular*}}
\label{tab4}
\end{table}

\begin{table}[ht]
\centering
\caption{Performance comparison between the CECT and SOTA models on the COVIDx CXR-3 dataset.}
\resizebox{0.84\linewidth}{!}{
\begin{tabular*}{504pt}{cccccccccc}
\toprule
    Category  & Model                                                  & Venue     & ACC             & NPV              & PPV             & SEN             & SPE             & FOS             \\
\midrule
    \multirow{8}{*}{Pure}   & ConvNeXt \citep{liu2022convnet}          & CVPR 22'  & 86.2\%          & 79.8\%           & 96.2\%          & 75.5\%          & 97.0\%          & 84.6\%          \\
                            & RegNetX \citep{radosavovic2020designing} & CVPR 20'  & 90.0\%          & 84.5\%           & 97.6\%          & 82.0\%          & 98.0\%          & 89.1\%          \\
                            & GhostNet \citep{han2020ghostnet}         & CVPR 20'  & 79.2\%          & 71.3\%           & 96.8\%          & 60.5\%          & 98.0\%          & 74.5\%          \\
                            & CSPNet \citep{wang2020cspnet}            & CVPRW 20' & 89.2\%          & 85.8\%           & 93.4\%          & 84.5\%          & 94.0\%          & 88.7\%          \\
                            & SKNet \citep{li2019selective}            & CVPR 19'  & 85.2\%          & 80.5\%           & 91.7\%          & 77.5\%          & 93.0\%          & 84.0\%          \\
                            & VoVNet \citep{lee2019energy}             & CVPRW 19' & 93.7\%          & 91.1\%           & 96.8\%          & 90.5\%          & 97.0\%          & 93.5\%          \\
                            & DeiT \citep{touvron2021training}         & ICML 21'  & 68.2\%          & 61.6\%           & 92.9\%          & 39.5\%          & 97.0\%          & 55.4\%          \\
                            & ConViT \citep{d2021convit}               & ICML 21'  & 79.2\%          & 73.5\%           & 88.7\%          & 67.0\%          & 91.5\%          & 76.4\%          \\
\midrule 
    \multirow{8}{*}{Hybrid} & MaxViT \citep{tu2022maxvit}              & ECCV 22'  & 74.0\%          & 66.8\%           & 92.1\%          & 52.5\%          & 95.5\%          & 66.9\%          \\
                            & CrossViT \citep{chen2021crossvit}        & ICCV 21'  & 63.5\%          & 60.5\%           & 68.7\%          & 49.5\%          & 77.5\%          & 57.6\%          \\
                            & LeViT \citep{graham2021levit}            & ICCV 21'  & 88.7\%          & 82.2\%           & 98.7\%          & 78.5\%          & \textbf{99.0\%} & 87.5\%          \\
                            & CoaT \citep{xu2021co}                    & ICCV 21'  & 81.7\%          & 74.3\%           & 95.7\%          & 66.5\%          & 97.0\%          & 78.5\%          \\
                            & MobileViT \citep{mehta2021mobilevit}     & ICLR 22'  & 75.2\%          & 70.6\%           & 82.6\%          & 64.0\%          & 86.5\%          & 72.1\%          \\
                            & CMT \citep{guo2022cmt}                   & CVPR 22'  & 84.0\%          & 77.6\%           & 94.2\%          & 72.5\%          & 95.5\%          & 81.9\%          \\
                            & MedViT \citep{manzari2023medvit}         & CIBM 23'  & 82.0\%          & 75.6\%           & 92.7\%          & 69.5\%          & 94.5\%          & 79.4\%          \\
                            & \textbf{CECT(Ours)}                      & -         & \textbf{97.2\%} & \textbf{96.1\%}  & \textbf{98.5\%} & \textbf{96.0\%} & 98.5\%          & \textbf{97.2\%} \\
\bottomrule
\end{tabular*}
}
\label{tab5}
\end{table}

We compare the model performance between our CECT and several SOTA models in \cref{tab4} and \cref{tab5}, for the COVID-19 radiography dataset and the COVIDx CXR-3 dataset, respectively. For intuitive illustration, we categorize the SOTA approaches into pure and hybrid methods, in which pure methods are CNN- or transformer-based, and hybrid methods are based on their integration. It can be found that CECT outperforms SOTA methods on all evaluation metrics to a different extent. From \cref{tab4}, we can find that CECT shows apparent performance leadership on most metrics. It reaches the highest ACC, NPV, SEN, and FOS of 98.1\%, 98.6\%, 96.1\%, and 96.4\%, respectively. Regarding the PPV and SPE, the performance is also among the best though not as high as GhostNet. From the result shown in \cref{tab5}, CECT demonstrates tremendous leadership on most evaluation metrics, reaching an ACC, NPV, PPV, SEN, and FOS of 97.2\%, 96.1\%, 98.5\%, 96.0\%, and 97.2\%, respectively. The highest SPE is achieved by LeViT with 0.5\% leadership compared with CECT. The overall performance leadership of CECT shows the effectiveness of the proposed architecture and the outstanding performance leadership demonstrates the importance of capturing both multi-local and global features.

\begin{figure*}
	\centering
	  \includegraphics[width=\textwidth]{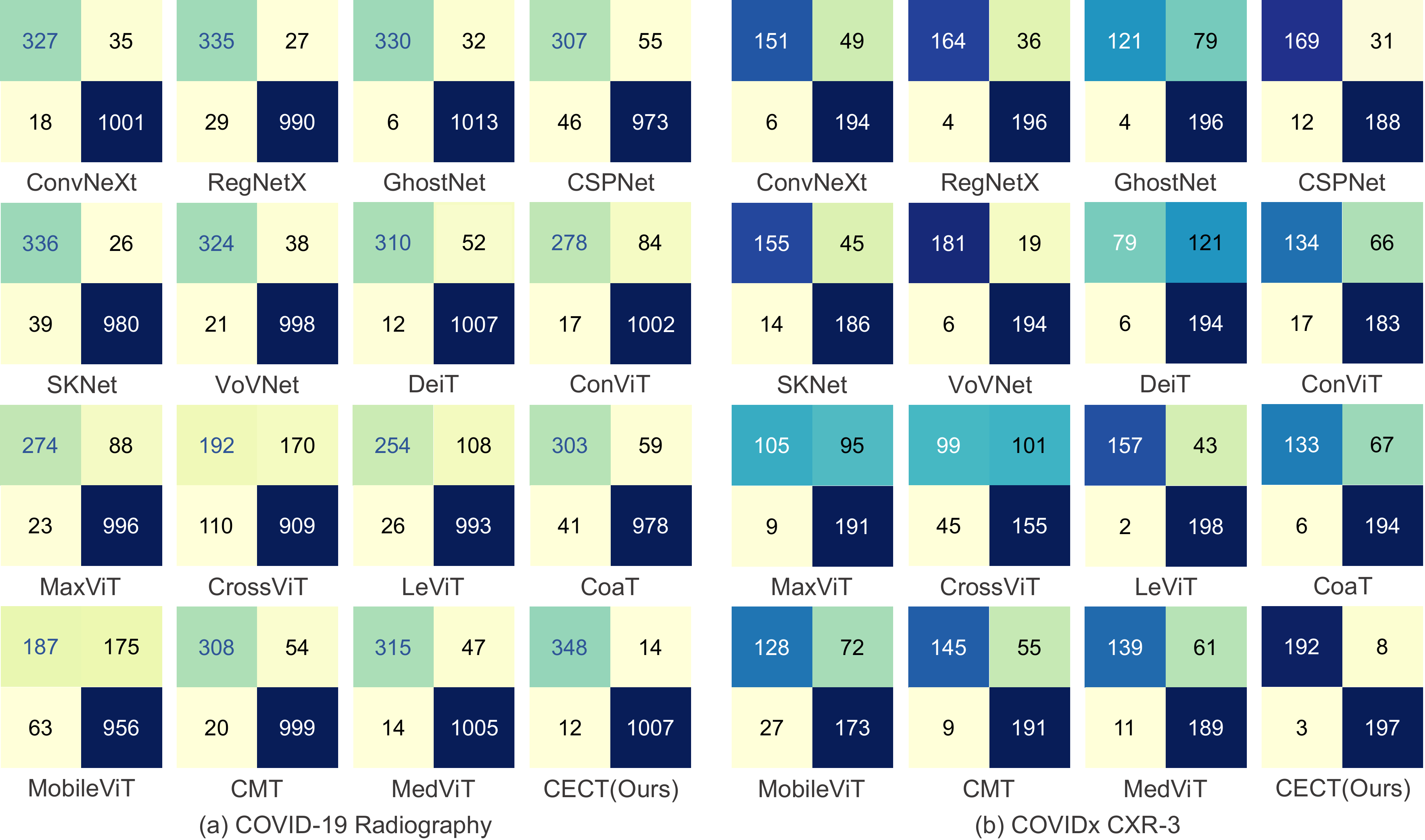}
	  \caption{The confusion matrix comparison between the CECT and SOTA models on both datasets. Four sub-blocks represent TP, FN, FP, and TN from left to right, top to bottom.}
	  \label{fig3}
\end{figure*}

We leverage the confusion matrix to visualize the model performance across CECT and SOTA methods in \cref{fig3}. It can be observed that the CECT outperforms existing SOTA methods on TP, TN, FP, and FN. For the COVID-19 radiography dataset, the proposed CECT misclassifies 26 images with 14 of them being FN and 12 being FP. Compared with SOTA methods, the CECT reaches the lowest FN and the second lowest FP. Though the GhostNet achieves the optimal FP, it does not perform well on FN and results in 38 misclassified images. Considering the COVIDx CXR-3 dataset, CECT miscategorized 11 images with 8 of them being FN and 3 being FP. Analogously, CECT realizes the lowest FN and the second lowest FP. Though LeViT demonstrates the lowest FP, it does not achieve an ideal FN and results in 45 misclassified images. This number of images is approximately four times compared with that of CECT. Poor performance is found for the DeiT and CrossViT as these approaches even result in a higher FN compared with TP. The superior performance of CECT alongside its ability to balance varying metrics highlights its effectiveness in COVID-19 image classification.

\begin{figure*}
	\centering
	  \includegraphics[width=\textwidth]{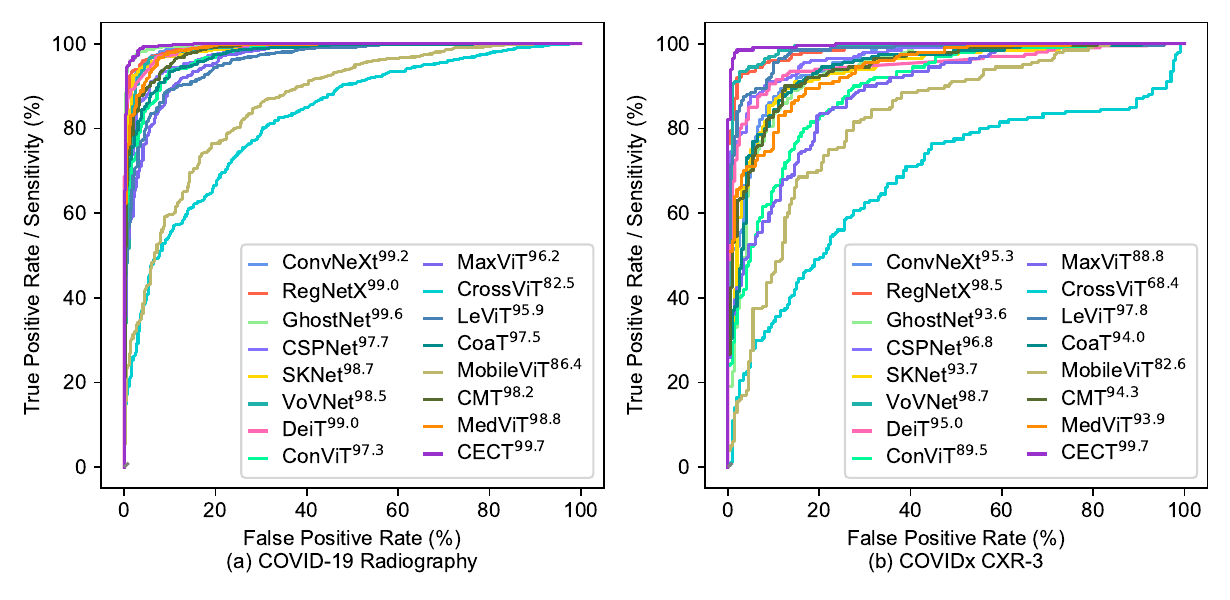}
	  \caption{Receiver operating characteristic curve across CECT and SOTA methods on the two datasets. Superscripts show the area under the curve in percentage. The false positive rate (FPR) presents the proportion of negative samples that are incorrectly predicted as positive. Specifically, $F P R=\frac{F P}{F P+T N}$.}
	  \label{fig4}
\end{figure*}

\begin{figure*}
	\centering
	  \includegraphics[width=\textwidth]{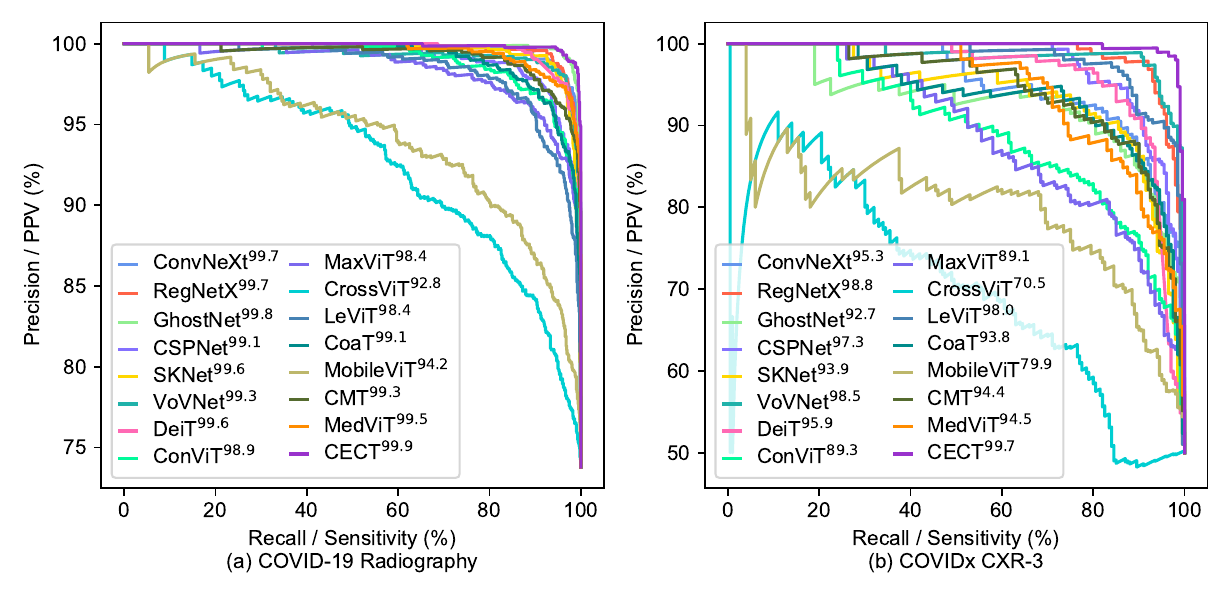}
	  \caption{Precision-recall curve across CECT and SOTA methods on the two datasets. Superscripts show the average precision in percentage.}
	  \label{fig5}
\end{figure*}

We visualize the receiver operating characteristic curve and precision-recall curve in \cref{fig4} and \cref{fig5} to further illustrate and compare the model performance. It can be found that CECT largely outperforms the SOTA methods on both datasets. From \cref{fig4}, it is observed that on the COVID-19 radiography dataset, CECT achieves the highest area under the curve (AUC) of 99.7\%, followed by the GhostNet. However, several SOTA methods including CrossViT and MobileViT result in a relatively low AUC of 82.5\% and 86.4\%, respectively. Regarding the COVIDx CXR-3 dataset, CECT demonstrates the highest AUC of 99.7\%, followed by the VoVNet with an AUC of 98.7\%. Though GhostNet achieves superior performance on the COVID-19 radiography dataset, it does not demonstrate outstanding performance under the current setup with an AUC of 93.6\%. Similarly, CrossViT shows a poor performance with an AUC of 68.4\%. In \cref{fig5}, analogous results are observed on the COVID-19 radiography dataset, in which the CECT reaches the highest average precision (AP) of 99.9\%, followed by GhostNet. The lowest performance is achieved by the CrossViT with an AP of 92.8\%. Regarding the COVIDx CXR-3 dataset, CECT overperforms the SOTA methods with an AP of 99.7\%. The second-highest AP of 98.8\% is achieved by RegNetX. The GhostNet that outperforms other methods on the COVID-19 radiography dataset achieves a relatively low AP of 92.7\% under the current configuration. Results from the two curves demonstrate the comprehensiveness and stability of the CECT.

\begin{figure*}
	\centering
	  \includegraphics[width=\textwidth]{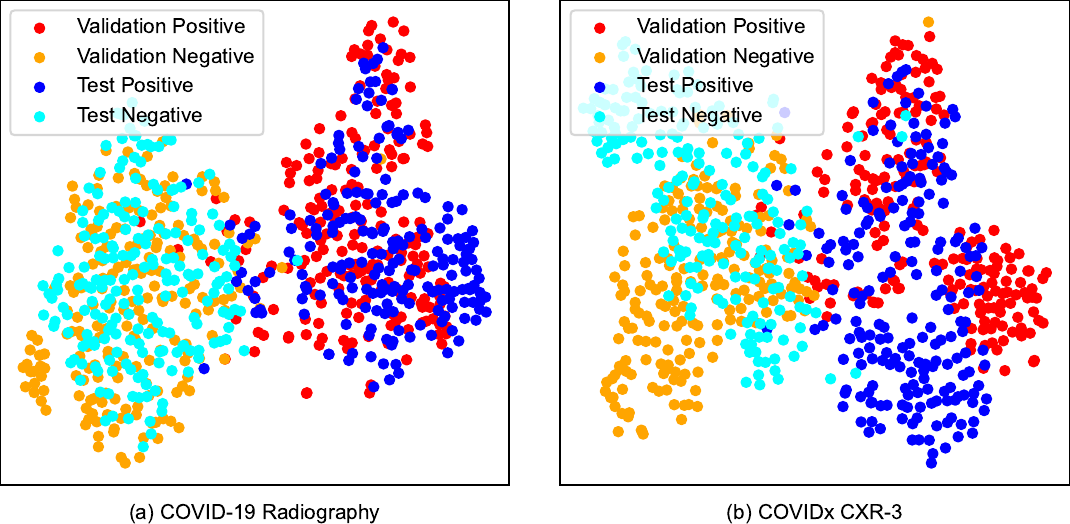}
	  \caption{The t-SNE visualization across the validation and test subsets on the two datasets.}
	  \label{fig6}
\end{figure*}

Upon examining \cref{tab4}, \cref{tab5} and associated results, a consistent and intriguing observation emerges. Specifically, the models overall perform worse in the COVIDx CXR-3 dataset, which has a larger size compared with the COVID-19 radiography dataset. This is contrary to our expectations as the performance of the model usually improves with increasing dataset size. To perform in-depth analysis for this observation, we leverage the t-distributed stochastic neighbor embedding (t-SNE) \cite{van2008visualizing} to visualize the data distribution across the two datasets in \cref{fig6}. We randomly sample 200 images for each category of each subset to prevent plethoric data points and the features are extracted ahead of the prediction head of the WAC. We compare the data distribution between the validation subset and the test subset and the reason is that the training subset undergoes data augmentations while the validation subset and test subset are not. Based on the observation, it becomes evident that the two subsets of the COVIDx CXR-3 dataset exhibit a relatively inconsistent data distribution, as evidenced by the significant non-overlapping regions between the red and blue data points. This discrepancy may be attributed to the manner of generating test subsets. Recall that the training, validation, and test subsets of the COVID-19 radiography dataset are divided simultaneously, while the COVIDx CXR-3 dataset has a separately provided test subset. The separately provided test subset may lead to relatively inconsistent distribution and thus reduce the model performance. Nevertheless, our CECT reaches an ideal performance on the COVIDx CXR-3 dataset, showing its remarkable generalization ability.

\begin{figure*}
	\centering
	  \includegraphics[width=\textwidth]{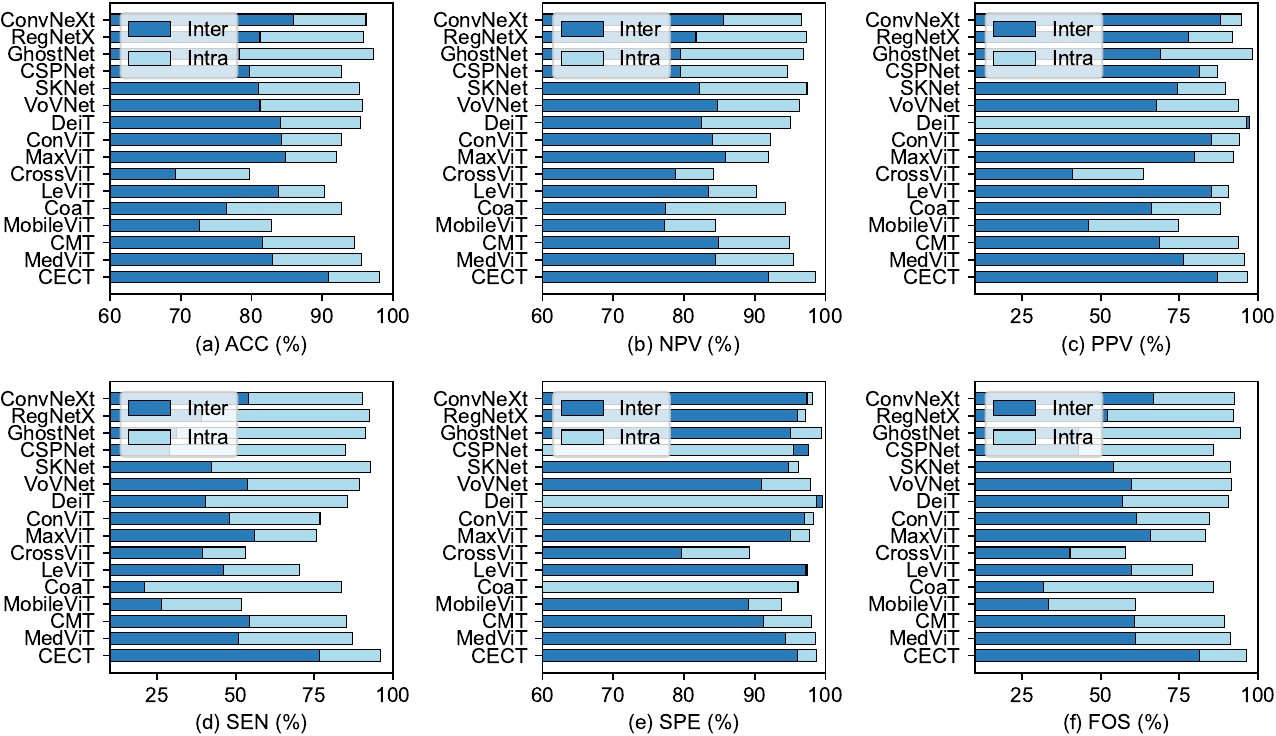}
	  \caption{Performance comparison across inter-dataset evaluation and intra-dataset evaluation. Models are trained on the COVIDx CXR-3 dataset while tested on the COVID-19 radiography dataset.}
	  \label{fig7}
\end{figure*}

To further demonstrate the generalization ability of CECT, we contrast its performance with SOTA methods on unseen datasets. Specifically, the models undergo training and testing on distinct datasets. Given the notable performance of most methods on the COVID-19 radiography dataset, we perform an inter-dataset evaluation using the COVIDx CXR-3 dataset for training and the COVID-19 radiography dataset for testing. The performance comparison across inter-dataset evaluation and intra-dataset evaluation can be found in \cref{fig7}. From the results, we can find that CECT outperforms SOTA methods to a large extent in the inter-dataset evaluation. It demonstrates superior performance among all metrics with an ACC of 90.9\%, showing a merely 7.2\% decrease compared with the intra-dataset evaluation. This is somewhat higher than the intra-dataset evaluation of several SOTA methods such as CrossViT and MobileViT. Considering NPV, SEN, and FOS, CECT achieves the highest results of 92.0\%, 76.5\%, and 81.5\%, respectively. The highest PPV and SPE are achieved by DeiT, demonstrating even higher results compared with the intra-dataset evaluation. This is unreasonable and therefore deserves questioning and further investigation. Upon close examination, it can be found that though DeiT shows exemplary performances on PPV and SPE, it severely underperformed on others such as SEN. Such results can suggest the over-prediction of certain scenarios and do not indicate superior performance. Similar results are observed for the CSPNet and CoaT. The extraordinary performance under such a challenging task demonstrates the generalization ability of the CECT.

\subsection{Ablation Study}
\label{4.3}

\begin{table}[ht]
\centering
\caption{Ablation study of CECT on the COVIDx CXR-3 dataset. The experiments are performed across different block configurations and feature capture scales.}
\resizebox{0.84\linewidth}{!}{
\begin{tabular*}{472pt}{ccccccccccc}
\toprule
    PCE         & ATD         & WAC         & 28 × 28     & 56 × 56    & 112 × 112  & 224 × 224  & $\alpha$    & $\beta$     & $\gamma$    & ACC \\
\midrule
    \checkmark  & $\times$    & $\times$    & \checkmark  & $\times$   & $\times$   & $\times$   & -           & -           & -           & 77.5\% \\ 
    \checkmark  & $\times$    & $\times$    & $\times$    & \checkmark & $\times$   & $\times$   & -           & -           & -           & 76.0\% \\ 
    \checkmark  & $\times$    & $\times$    & $\times$    & $\times$   & \checkmark & $\times$   & -           & -           & -           & 85.2\% \\ 
    $\times$    & $\times$    & \checkmark  & $\times$    & $\times$   & $\times$   & \checkmark & -           & -           & -           & 85.0\% \\ 
    \checkmark  & \checkmark  & \checkmark  & $\times$    & $\times$   & \checkmark & \checkmark & 0           & 0           & 1           & 74.7\% \\ 
    \checkmark  & \checkmark  & \checkmark  & $\times$    & \checkmark & \checkmark & \checkmark & 0           & 0.5         & 0.5         & 84.2\% \\ 
    \checkmark  & \checkmark  & \checkmark  & \checkmark  & \checkmark & \checkmark & \checkmark & $0.\dot{3}$ & $0.\dot{3}$ & $0.\dot{3}$ & 97.2\% \\ 
\bottomrule
\end{tabular*}
}
\label{tab7}
\end{table}

\begin{figure*}
	\centering
	  \includegraphics[width=\textwidth]{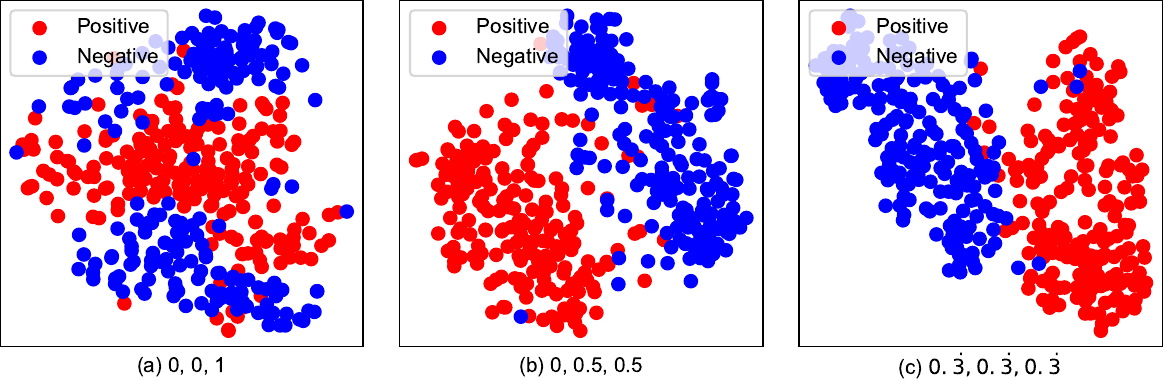}
	  \caption{The t-SNE visualization across various groups of ensemble coefficients on the COVIDx CXR-3 dataset.}
	  \label{fig8}
\end{figure*}

To demonstrate the effectiveness and importance of capturing both multi-local and global features, we perform extensive ablation experiments on the COVIDx CXR-3 dataset as CECT outperforms SOTA methods to a large extent. The experiments are performed across different block configurations and feature capture scales and the results can be found in \cref{tab7}. For the variants with PCE purely, we assess the performance of the three sub-encoders with designed classification heads. For the variant merely consisting of WAC, we train it with the prediction head. The ATD is not evaluated separately as it serves for decoding. In case all PCE, ATD, and WAC exist, we simulate the scenarios in which local features at varying scales are omitted. This can result in two variants, in which the one lacks local features at both 28 × 28 and 56 × 56 scales and the other lacks at the 28 × 28 scale only. It is clear that the variants employing PCE or WAC are designed to simulate cases utilizing either CNN- or transformer-based architecture. Conversely, variants including PCE, ATD, and WAC simulate scenarios where CNN- and transformer-based architectures are married. Upon examination, we note a substantial decrease in model performance when either the CNN- or transformer-based blocks are used exclusively. We observe the highest accuracy of 85.2\% and 85.0\%, for the variants utilizing CNN- and transformer-based architectures, respectively. When integrating both CNN- and transformer-based methods, it is observed that the overall performance deteriorates as more features are absent. When the model lacks local features at the 28 × 28 scale, the observed accuracy stands at 84.2\%. With the absence of local features at both the 28 × 28 and 56 × 56 scales, the performance further declines to 74.7\%. We present the t-SNE visualization across different groups of ensemble coefficients in \cref{fig8} to illustrate the results intuitively. From the results, it can be inferred that as the amount of features captured increases, the discriminative ability of the model improves noticeably. This can further underscore the effectiveness and importance of capturing multi-scale features from the input, thereby highlighting the novelty of our CECT approach.


\section{Conclusion}
\label{5}

In this paper, we propose a novel CECT model by controllable ensemble CNN and transformer for COVID-19 classification. The CECT can extract features at both multi-local and global scales without sophisticated module design. Moreover, the contribution of local features at different scales can be arbitrarily controlled with the proposed ensemble coefficients. Extensive intra-dataset and inter-dataset experiments on two public COVID-19 datasets demonstrate that CECT surpasses existing methods, whether pure CNN- or transformer-based or their integration. The extraordinary performance and generalization ability demonstrate the effectiveness of the proposed CECT. To reveal the effectiveness and importance of capturing both multi-local and global features, we perform extensive ablation experiments on the COVIDx CXR-3 dataset and the results show that features at each scale count. The efficacy of CECT underscores the notion that increasing the complexity of the network architecture is not invariably essential for enhancing performance. A streamlined yet effective architecture can not only achieve superior performance but also be spread and applied more easily. While CECT exhibits outstanding performance in comparison to SOTA methods, it is pertinent to highlight that its parameter count can be relatively large. This increased computational demand stems from the integration of various blocks and branches, potentially making it less suitable for mobile applications. The future perspectives of CECT can be two-fold. Firstly, given that the CECT is tailored for image classification, it can be valuable to adapt it for fine-grained tasks, notably image segmentation. Unlike classification, segmentation aims at capturing pixel-level features from the input and has a more strict requirement for capturing features at different scales due to the various sizes across different objects. Under this scenario, the CECT-based architecture can further distinguish its strengths. Secondly, the selection of the optimal coefficient group could be made more intrinsic and adaptive, rather than being pre-determined. For instance, the coefficients can be integrated as adaptive hyperparameters, updated iteratively during the training process to optimize model performance. This approach could eliminate the need for separate coefficient searching and offer a greater variety of combinations.



\bibliographystyle{unsrt}
\bibliography{reference.bib}

\biboptions{sort&compress}







\end{sloppypar}
\end{document}